\documentclass[10pt,conference]{IEEEtran}
\usepackage{multirow}
\usepackage{graphicx}
\usepackage{siunitx}
\usepackage{relsize}
\usepackage{amsfonts}
\usepackage{amssymb}
\usepackage{amsmath}
\usepackage{verbatim}
\usepackage{amssymb}
\usepackage{mathrsfs}
\usepackage{epsfig}
\usepackage{graphicx}
\usepackage{amsthm}
\usepackage{balance}
\usepackage{bbm}
\usepackage{bm}
\usepackage{cite}
\usepackage{txfonts}
\usepackage{array}
\usepackage{tabularx, booktabs}
\usepackage{comment}
\usepackage[caption=false]{subfig}

\graphicspath{{./Figure/}}
\begin{document}

\bstctlcite{IEEEexample:BSTcontrol}
\pagestyle{empty}
\title{Low-Complexity Pilot-Aided Doppler Ambiguity Estimation for OTFS Parametric Channel Estimation}
\author{Bo-Yuan Chen and Hsuan-Jung Su\\
Graduate Institute of Communication Engineering\\
National Taiwan University\\
Taipei, Taiwan\\
Email:  \{r13942114, hjs\}@ntu.edu.tw}
\maketitle
\thispagestyle{empty}

\begin{abstract} Orthogonal Time Frequency Space (OTFS) modulation offers robust performance in high-mobility scenarios by transforming time-varying channels into the delay-Doppler (DD) domain.
However, in high-mobility environment such as emerging 5G Non-Terrestrial Networks (NTN), the extreme orbital velocities of Low Earth Orbit (LEO) satellites frequently cause the physical Doppler shifts to exceed the fundamental grid range.
This Doppler ambiguity induces severe model mismatch and renders traditional MLE channel estimators ineffective.
To address this challenge, this paper proposes a novel low-complexity pilot-aided Doppler ambiguity detection and compensation framework.
We first mathematically derive the OTFS input-output relationship in the presence of aliasing, revealing that Doppler ambiguity manifests itself as a distinct phase rotation along the delay dimension.
Leveraging this insight, we developed a two-stage estimator that utilizes pairwise phase differences between pilot symbols to identify the integer ambiguity, followed by a refined Maximum Likelihood Estimation (MLE) for channel recovery.
We investigate two pilot arrangements, Embedded Pilot with Guard Zone (EP-GZ) and Data-Surrounded Pilot (DSP), to analyze the trade-off between interference suppression and spectral efficiency.
Simulation results demonstrate that the proposed scheme effectively eliminates the error floor caused by ambiguity, achieving Bit Error Rate (BER) and Normalized Mean Square Error (NMSE) performance comparable to the exhaustive search benchmark while maintaining a computational complexity similar to standard MLE. \end{abstract}


\section{Introduction}\label{introduction}
\IEEEPARstart{N}{ext}-generation wireless communication services, such as for high-speed trains (HST) or Low Earth Orbit (LEO) satellite networks, are characterized by high-mobility scenarios where large Doppler shifts pose significant challenges to link reliability. Traditional Orthogonal Frequency Division Multiplexing (OFDM) suffers severely from inter-carrier interference (ICI) in such environments, leading to performance degradation. To address this, Orthogonal Time Frequency Space (OTFS) modulation has emerged as a promising waveform \cite{hadani2017orthogonal}. By modulating information symbols in the delay-Doppler (DD) domain, OTFS transforms the time-varying fading channel into a quasi-static interaction, providing robust performance in high-Doppler channels \cite{raviteja2018interference}\cite{wei2021otfs}.

Realizing the potential of OTFS requires an accurate channel estimation. Although sparse recovery algorithms such as orthogonal matching pursuit (OMP) \cite{rasheed2020sparse} and sparse Bayesian learning (SBL) \cite{wei2022offgrid} are effective, they incur high computational costs due to large-scale matrix inversions. To address this, low-complexity parametric approaches such as Modified Maximum Likelihood Estimation (MLE) \cite{khan2023low}\cite{muppaneni2023channel} have been proposed. Unlike direct channel matrix estimation, these methods explicitly estimate key path parameters (i.e., gains, delays, and Doppler shifts) to reconstruct the effective channel matrix. This strategy avoids the computationally expensive matrix inversions inherent in OMP and SBL, offering a significant reduction in complexity while maintaining accuracy.

However, a critical limitation of these schemes is the assumption that the normalized Doppler shift is strictly confined within the fundamental range $[-N/2, N/2]$.
This assumption is readily violated in extremely high-mobility environment, such as 5G NTN.
Due to the extremely high orbital velocities of LEO satellites, the resulting Doppler shift frequently exceeds the unambiguous limit.
Under these conditions, the physical Doppler shift spills over the grid boundaries, leading to severe model mismatch that renders standard estimators (e.g., conventional MLE) ineffective.
Although this ambiguity could theoretically be resolved through an exhaustive search (referred to as \textit{Extended MLE}), such a brute-force approach suffers from prohibitive computational complexity, making it impractical for real-time applications.
Consequently, there is a critical need for a low-complexity solution that can robustly detect and compensate for Doppler ambiguity.

In this paper, we address this gap by proposing a novel pilot-aided Doppler ambiguity detection and compensation framework. We first derive the input-output relationship of the OTFS system in the presence of Doppler ambiguity, explicitly characterizing the phase rotation induced by the aliasing effect. Based on this derivation, we develop a two-stage estimator that first estimates the integer ambiguity using a multi-pilot structure and subsequently performs channel estimation on the compensated signal. We investigate two pilot arrangements: the Embedded Pilot with Guard Zone (EP-GZ) and the Data-Superimposed Pilot (DSP), to explore the trade-off between interference suppression and spectral efficiency.

The main contributions of this paper are summarized as follows:
\begin{itemize}
  \item We analyze the impact of high-mobility Doppler shifts on OTFS systems and mathematically derive the phase rotation effect caused by Doppler ambiguity. This derivation provides the theoretical basis for the estimation of ambiguity.
  \item We propose a novel pilot-aided ambiguity estimation and compensation scheme. Using the phase difference between pairwise pilot symbols, the proposed method can accurately identify the integer ambiguity before estimating the channel.
  \item We evaluate the proposed estimator using two different pilot arrangements. Simulation results verify that our method effectively eliminates the error floor caused by ambiguity, achieving BER and NMSE performance close to the Perfect channel state information (CSI) case and the high-complexity Extended MLE benchmark.
\end{itemize}

\textit{Notations}: Vectors and matrices are denoted by boldface lower-case and upper-case letters, respectively. $(\cdot)^T$, $(\cdot)^*$, and $(\cdot)^H$ denote transpose, conjugate, and conjugate transpose operations, respectively. $\mathbf{I}_N$ is the identity matrix of dimension $N$. $\mathbf{e}_i^{(D)}$ denotes the standard basis vector of dimension $D$ with a 1 at index $i$ and 0 elsewhere. $\mathbf{\Pi}_{M,m} = [\mathbf{e}_{m+1}^{(M)}, \dots, \mathbf{e}_{M}^{(M)}, \mathbf{e}_{1}^{(M)}, \dots, \mathbf{e}_{m}^{(M)}]$ denotes the $m$-step forward cyclic shift permutation matrix of dimension $M$. $[\mathbf{A}]_{m,n}$ represents the element on the $m$-th row and $n$-th column of matrix $\mathbf{A}$. $\odot$ denotes the Hadamard product. $\langle \cdot \rangle_M$ denotes the modulo-$M$ operation.

\section{System Model}\label{sec:system_model}
Consider an OTFS modulation system operating on a DD grid $\Gamma$, which consists of $M$ delay bins and $N$ Doppler bins.
The DD grid is defined as $\Gamma = \{ (k, l) \mid k = 0, \dots, N-1, l = 0, \dots, M-1 \}$, where $k$ and $l$ denote the Doppler and delay indices, respectively.
The fundamental Doppler resolution is $\Delta\nu = 1/(NT)$, and the delay resolution is $\Delta\tau = 1/(M\Delta f)$, where $T$ is the OTFS symbol duration and $\Delta f$ is the subcarrier spacing (SCS).

    \subsection{SFFT-based OTFS Architecture}\label{subsec:otfs_arch}
    In this paper, we adopt the standard \textit{Symplectic Finite Fourier Transform} (SFFT) based OTFS architecture described in \cite{hadani2017orthogonal}.
    The DD domain transmit symbol $x_{\mathrm{DD}}[k,l]$ denotes the symbol placed at the $k$-th Doppler and the $l$-th delay bin.
    At the transmitter, these symbols are mapped to the Time-Frequency (TF) domain via the \textit{Inverse Symplectic Finite Fourier Transform} (ISFFT) and subsequently converted into a continuous-time signal $s(t)$ using the \textit{Heisenberg transform}.
    The signal propagates through a wireless channel and is received as $r(t)$.
    At the receiver, the \textit{Wigner transform} and SFFT are applied to convert the received signal back to the DD domain symbols $y_{\mathrm{DD}}[k,l]$ for channel estimation and data detection.

    The wireless channel in a high-mobility environment can be modeled as a sparse multipath channel in the DD domain.
    The channel impulse response with $P$ propagation paths is given by:
    \begin{equation}\label{eq:channel_impulse_response}
        h(\tau, \nu) = \sum_{i=1}^{P} h_i \delta(\tau - \tau_i) \delta(\nu - \nu_i),
    \end{equation}
    where $h_i$, $\tau_i$, and $\nu_i$ denote the complex path gain, delay, and Doppler shift associated with the $i$-th path, respectively.

    These physical parameters are related to the normalized DD grid indices as follows:
    \begin{equation}\label{eq:delay_doppler_def}
        \tau_i = \frac{l_{i}}{M \Delta f}, \quad \nu_i = \frac{k_{i}}{NT}.
    \end{equation}
    here, $l_{i} \in \mathbb{Z}$ is the normalized delay index. We assume that fractional delays are negligible given the fine time resolution $1/(M\Delta f)$ in wideband systems \cite{tse2005fundamentals}.
    For the Doppler dimension, the normalized Doppler shift $k_{i} \in \mathbb{R}$ can be decomposed as $k_{i} = K_{i} + \kappa_{i}$, where $K_{i} \in \mathbb{Z}$ is the integer Doppler index and $\kappa_{i} \in [-0.5, 0.5]$ represents the fractional Doppler shift.

    \subsection{DD Domain Input-Output Relation}\label{subsec:io_relation}
    To facilitate the analysis of Doppler ambiguity, we adopt the \textit{fast/slow-time matrix representation} derived in \cite{xia2024achieving}.
    By arranging the DD domain transmit symbols into a matrix $\mathbf{X}_{\mathrm{DD}} \in \mathbb{C}^{M \times N}$ and reformulating the input-output (I/O) relation in \cite[Eq. (14)]{xia2024achieving} using the normalized parameters defined in \eqref{eq:delay_doppler_def}, the DD domain matrix I/O relation can be expressed as:
    \begin{equation}\label{eq:io_matrix}
        \mathbf{Y}_{\mathrm{DD}} = \sum_{i=1}^{P} h_i
        \mathbf{Q}_{\mathrm{ISI}}^{\mathrm{Fast}}(l_i)
        \mathbf{D}(k_i) \tilde{\mathbf{X}}_{\mathrm{DD}}(l_i,k_i) \mathbf{Q}_{\mathrm{ISI}}^{\mathrm{Slow}}(k_i)
        + \mathbf{W}_{\mathrm{DD}},
    \end{equation}
    where $\mathbf{W}_{\mathrm{DD}} \in \mathbb{C}^{M \times N}$ is the effective additive white Gaussian noise (AWGN) matrix in the DD domain.
    The matrices $\mathbf{Q}_{\mathrm{ISI}}^{\mathrm{Fast}}(l_i) \in \mathbb{C}^{M \times M}$ and $\mathbf{Q}_{\mathrm{ISI}}^{\mathrm{Slow}}(k_i) \in \mathbb{C}^{N \times N}$ capture the circular shifts and spreading effects along the fats-time (delay) and slow-time (Doppler) dimensions, respectively.
    The diagonal matrix $\mathbf{D}(k_i) \in \mathbb{C}^{M \times M}$ accounts for the phase rotation along the fast-time dimension.
    $\tilde{\mathbf{X}}_{\mathrm{DD}}(l_i,k_i) \in \mathbb{C}^{M \times N}$ incorporates transmit symbols and pulse-shaping effects.
    Finally, $\mathbf{Y}_{\mathrm{DD}}\in \mathbb{C}^{M \times N}$ is the matrix form of received signal in the DD domain.
    Detailed definitions of these matrices can be found in \cite{xia2024achieving}.

\section{Doppler Ambiguity Analysis}\label{sec:DA_analysis}
In this section, we analyze the physical conditions that give rise to Doppler ambiguity and its impact on the DD domain I/O relation.

    \subsection{Physical Constraints on Unambiguous Doppler}\label{subsec:problem_formulation}

    In the vast majority of existing OTFS literature, a fundamental assumption is that the Doppler shift imposed by the channel is strictly confined within the fundamental period of the Doppler grid.
    Specifically, based on the normalized Doppler shift $k_i$ defined in \eqref{eq:delay_doppler_def}, the standard assumption requires:
    \begin{equation}\label{eq:standard_assumption}
        -\frac{N}{2} < k_i < \frac{N}{2}.
    \end{equation}
    This assumption guarantees a unique, one-to-one mapping between the physical Doppler frequency and the indices on the DD grid (i.e., $|K_i| < N/2$), which is a critical prerequisite for the success of conventional parametric channel estimation algorithms.

    However, as wireless systems evolve, this assumption faces physical challenges.
    The normalized Doppler shift is determined by the relative velocity $v_i$ between the transmitter and the receiver, related by $k_i = \frac{v_i f_c}{c} NT$, where $f_c$ is the carrier frequency and $c$ is the speed of light.
    Substituting this into \eqref{eq:standard_assumption}, the condition for an unambiguous Doppler representation implies a strict upper bound on the relative velocity:
    \begin{equation}\label{eq:vel_constraint}
        |v_i| < \frac{c}{2} \cdot \frac{\Delta f}{f_c}.
    \end{equation}

    This fundamental assumption becomes invalid as wireless systems evolve to support high-mobility scenarios, particularly in 5G NTN operating in the L-/S-bands.
    While LEO satellites are essential for global coverage, they orbit at extremely high velocities that far exceed the unambiguous limit defined in \eqref{eq:vel_constraint}.
    To mitigate this, one might intuitively increase the SCS to widen the unambiguous range.
    However, adjusting the SCS only affects the degree of the ambiguity but does not eliminate it.\footnote{To illustrate the severity, consider a standard 5G NTN configuration in the S-band ($f_c = 2$ GHz) defined in \cite{3gpp_tr38741_v1860}. Substituting the subcarrier spacings $\Delta f \in \{15, 30, 60\}$ kHz (corresponding to numerologies $\mu=0, 1, 2$) into \eqref{eq:vel_constraint} yields maximum unambiguous velocities of $v_{\max} \approx 4,050$ km/h, $8,100$ km/h, and $16,200$ km/h, respectively. However, the maximum Doppler shift of a LEO satellite at $600$ km altitude is approximately $46$ kHz \cite{tuninato20245gntn}, corresponding to a relative velocity of roughly $24,800$ km/h. This reveals that even with the largest supported SCS ($\mu=2$), the satellite velocity still exceeds the unambiguous limit. Crucially, in bandwidth-limited scenarios (e.g., 5 MHz channel) where a small SCS ($\mu=0$) is often mandatory to accommodate the SSB, the actual velocity is more than \textit{6 times} the unambiguous limit. This confirms that Doppler ambiguity is unavoidable in practical NTN scenarios.}
    As detailed in \cite{3gpp_tr38741_v1860}, even the largest supported SCS for S-band operations is insufficient to accommodate the massive Doppler shifts induced by LEO orbital velocities.
    Consequently, under high-mobility satellite conditions, the Doppler shift inevitably exceeds the grid boundaries ($|k_i| \ge N/2$), leading to \textit{Doppler ambiguity}.

    \subsection{Mathematical Modeling of Doppler Ambiguity}\label{subsec:da_modeling}

    Given the velocity constraint derived in \eqref{eq:vel_constraint}, we now formulate the mathematical impact of Doppler ambiguity on the DD domain input-output relation.

    Let $k_i$ denote the \textit{true normalized Doppler shift} of the $i$-th path as defined in \eqref{eq:delay_doppler_def}.
    When the relative velocity exceeds the unambiguous limit, $k_i$ exceeds the range of the fundamental Doppler grid.
    We naturally decompose the true Doppler shift $k_i$ into a base component and an integer ambiguity component:
    \begin{equation}\label{eq:doppler_decomp}
        k_i = \tilde{k}_i + N_{\mathrm{amb}, i} N,
    \end{equation}
    where $\tilde{k}_i=\tilde{K}_i+\tilde{\kappa}_i \in (-N/2, N/2)$ represents the \textit{base normalized Doppler shift} observable on the grid, and $N_{\mathrm{amb}, i} \in \mathbb{Z}$ represents the \textit{Doppler ambiguity integer} (i.e., the number of wrapped grid periods).
    Substituting the decomposition from \eqref{eq:doppler_decomp} into the matrix I/O relation \eqref{eq:io_matrix}, we examine the impact of the ambiguity integer $N_{\mathrm{amb}, i}$ on the Doppler-dependent matrices.

    First, consider the slow-time interaction matrix $\mathbf{Q}_{\mathrm{ISI}}^{\mathrm{Slow}}(k_i)$ and the effective symbol matrix $\tilde{\mathbf{X}}_{\mathrm{DD}}(l_i, k_i)$. These matrices capture the Doppler-induced phase rotations across the slow-time dimension, which are determined by the normalized frequency factor $\frac{k_i}{N}$. Consequently, both matrices exhibit periodicity of $N$ along the Doppler dimension.
    Substituting \eqref{eq:doppler_decomp} into the phase term $e^{-j2\pi \frac{k_i}{N}}$, we obtain:
    \begin{equation}
        e^{-j2\pi \frac{k_i}{N}}
        = e^{-j2\pi \frac{\tilde{k}_i + N_{\mathrm{amb}, i}}{N}}
        = e^{-j2\pi \frac{\tilde{k}_i}{N}} \times e^{-j2\pi N_{\mathrm{amb}, i}}.
    \end{equation}
    Since $N_{\mathrm{amb}, i}$ is an integer, the term $e^{-j2\pi N_{\mathrm{amb}, i}} = 1$.
    This implies that the integer ambiguity component is invisible in these matrices, making them indistinguishable from their base Doppler counterparts.

    Next, consider the fast-time phase rotation matrix $\mathbf{D}(k_i)$, which operates on the sampling interval $T/M$.
    The phase rotation for a normalized Doppler shift $k_i$ at the sampling index $m$ is determined by the factor $e^{j2\pi \frac{k_i m}{NM}}$.
    Substituting the decomposition \eqref{eq:doppler_decomp} into this exponent, the $m$-th diagonal element becomes:
     \begin{equation}\label{eq:phase_ambiguity}
        [\mathbf{D}(k_i)]_{m,m} = e^{j2\pi \frac{(\tilde{k}_i + N_{\mathrm{amb}, i} N) m}{NM}} = \underbrace{e^{j2\pi \frac{\tilde{k}_i m}{NM}}}_{\text{Base Phase}} \times \underbrace{e^{j2\pi N_{\mathrm{amb}, i} \frac{m}{M}}}_{\text{Ambiguity Phase}}
    \end{equation}
    Unlike the slow-time dimension, Equation \eqref{eq:phase_ambiguity} shows that Doppler ambiguity introduces a residual phase rotation $e^{j2\pi N_{\mathrm{amb}, i} \frac{m}{M}}$ along the fast-time dimension.
    The effective matrix in the presence of ambiguity can be expressed as:
    \begin{equation}\label{eq:io_matrix_amb}
        \mathbf{D}(k_i) = \mathbf{D}(\tilde{k}_i) \mathbf{\Phi}(N_{\mathrm{amb}, i}),
    \end{equation}
    where $\mathbf{\Phi}(N_{\mathrm{amb}, i}) = \mathrm{diag}\{1, e^{j2\pi \frac{N_{\mathrm{amb}, i}}{M}}, \dots, e^{j2\pi \frac{N_{\mathrm{amb}, i}(M-1)}{M}}\}$ represents the phase matrix induced by Doppler ambiguity.

    Ideally, in the absence of Doppler ambiguity ($N_{\mathrm{amb}, i} = 0$), the phase matrix $\mathbf{\Phi}(N_{\mathrm{amb}, i})$ reduces to the identity matrix $\mathbf{I}_M$.
    In this case, \eqref{eq:io_matrix_amb} simplifies to $\mathbf{D}(k_i) = \mathbf{D}(\tilde{k}_i)$, consistent with the conventional system model where the physical Doppler is assumed to be within the fundamental grid limit.

    In contrast, when $N_{\mathrm{amb}, i} \neq 0$, the uncompensated row-wise phase rotation $\mathbf{\Phi}(N_{\mathrm{amb}, i})$ introduces a significant model mismatch.
    Conventional MLE-based channel estimation methods \cite{khan2023low}, which typically ignore this phase induced by ambiguity, fail to capture the true channel information.
    This estimation error propagates to the data detection stage, leading to severe performance degradation, as will be demonstrated in Section~\ref{simulation_result}.
    Although a brute-force search over all possible Doppler ambiguity integers (denoted as Extended MLE) could theoretically resolve this mismatch, it incurs a prohibitive computational cost, making it impractical for real-time applications.

\section{Proposed Pilot-Aided Doppler Ambiguity Estimation and Channel Estimation}\label{sec:proposed_method}

Equation \eqref{eq:io_matrix_amb} indicates that the ambiguity integer $N_{\mathrm{amb}, i}$ is located in the phase variation along the delay dimension. Leveraging this insight, we propose a pilot-aided iterative detection framework combining MLE with successive interference cancelation (SIC). For each propagation path, the proposed estimator sequentially performs: 1) \textbf{coarse pilot localization} via noncoherent energy accumulation; 2) \textbf{Doppler ambiguity estimation} using the identified pilot phases; 3) \textbf{phase compensation} on the received signal; 4) \textbf{refined MLE} for fractional Doppler and gain recovery; and 5) \textbf{SIC} to facilitate the detection of subsequent paths.

    \subsection{Pilot Arrangement Strategies}
    \label{subsec:pilot_arrangement}

    To enable pilot-aided estimation, we first decompose the DD domain transmit symbols $\mathbf{X}_{\mathrm{DD}}$ into pilot components $\mathbf{X}_{\mathrm{p}}\in \mathbb{R}^{M \times N}$ and data components $\mathbf{X}_{\mathrm{d}}\in \mathbb{C}^{M \times N}$:
    \begin{equation}
        \mathbf{X}_{\mathrm{DD}}
        = \mathbf{X}_{\mathrm{p}} + \mathbf{X}_{\mathrm{d}}
        = \sum_{u=1}^{N_{\mathrm{p}}} x_{p,u} \mathbf{e}_{l_{p,u}}^{(M)} (\mathbf{e}_{k_{p,u}}^{(N)})^T + \mathbf{X}_{\mathrm{d}},
    \end{equation}
    where $N_{\mathrm{p}}$ denotes the total number of pilot symbols. For the $u$-th pilot, $x_{p,u}\in\mathbb{R}$ is the known pilot symbol placed at $(l_{p,u}, k_{p,u})$.

\begin{figure}[!t]
        \graphicspath{{./Figure/}}
        \centering
        \includegraphics[scale=0.26]{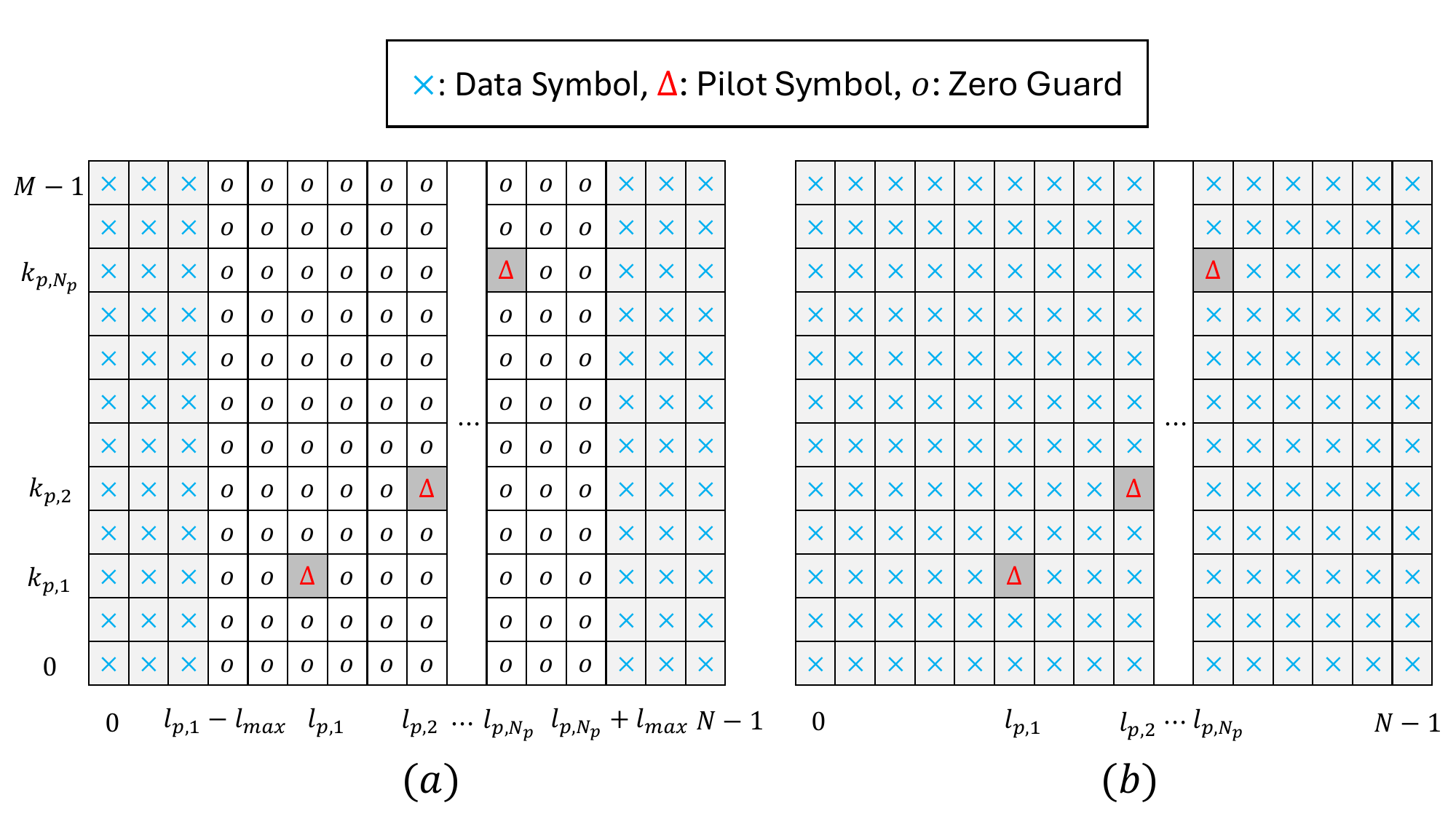}
        \caption{Illustration of the proposed pilot arrangement strategies in the DD domain: (a) Embedded Pilot with Guard Zone (EP-GZ); (b) Data-Surrounded Pilot (DSP).}
        \label{pilot_arrangement}
    \end{figure}

    We investigate two strategies for pilot and data symbol placement, balancing the trade-off between sensing accuracy and spectral efficiency. Adapted from \cite{raviteja2019embedded}, the EP-GZ scheme inserts zero guard region before and after each pilot to prevent inter-symbol interference (ISI) caused by channel spreading. Although this guarantees a reliable estimate, it incurs additional spectral overhead. In contrast, the DSP scheme removes the guard zones entirely to maximize spectral efficiency. Consequently, the pilots are inevitably exposed to interference leakage from neighboring data symbols. The proposed strategies are visually illustrated in Fig. \ref{pilot_arrangement}.

    \subsection{Received Pilot Signal Model and Phase Analysis}
    \label{subsec:rx_pilot_model}

    To theoretically justify the proposed ambiguity estimation scheme, we derive the explicit expression of the received pilot signal. Without loss of generality, we assume that the $i$-th path corresponds to the strongest signal component in the $i$-th iteration. Based on the decomposition in \eqref{eq:io_matrix}, we focus on the response of the $u$-th pilot symbol $x_{p,u}$. The component of the received signal corresponding to this specific pilot and path, denoted as $\mathbf{Y}_{\mathrm{p,u}}^{(i)}$, can be analyzed by following the transformation of the pilot basis vectors through the channel matrices. In this derivation, contributions from other path components and data symbols are considered as interference and are omitted to facilitate the analysis.

    First, the Doppler-induced phase rotation is governed by the fast-time matrix $\mathbf{D}(k_i)$. Applying this matrix to the effective pilot symbol $\tilde{\mathbf{X}}_{p,u}$ yields a phase shift dependent on the pilot's delay index:
    \begin{align}\label{eq:phase_derivation}
        \mathbf{D}(k_i) \tilde{\mathbf{X}}_{p,u}(l_i,k_i) &= \mathbf{D}(k_i) \left( x_{p,u} \mathbf{e}_{l_{p,u}}^{(M)} (\mathbf{e}_{k_{p,u}}^{(N)})^T \right) \nonumber\odot\bm{\Lambda}(l_i,k_i) \\
        &= x_{p,u} \lambda_{i,u} e^{j\phi(l_{p,u}, k_i)} \mathbf{e}_{l_{p,u}}^{(M)} (\mathbf{e}_{k_{p,u}}^{(N)})^T,
    \end{align}
    where $\lambda_{i,u} = [\bm{\Lambda}(l_i, k_i)]_{l_{p,u}, k_{p,u}}$ represents the boundary scaling coefficient{\footnote{By strictly enforcing the delay constraint $l_{p,u} < M - l_{\max} - 1$ for all pilots (i.e., $\forall u \in \{1, \dots, N_{\mathrm{pilot}}\}$), we ensure that the pulses received from all multipath components are fully contained within the observation window, guaranteeing that no boundary truncation occurs, thus satisfying $\lambda_{i,u} = 1, \forall u, i$.}}.

    The critical phase term $\phi(l_{p,u}, k_i)$ is derived as follows. By substituting the true Doppler decomposition $k_i = \tilde{k}_i + N_{\mathrm{amb}, i}N$, we observe that unlike the slow-time interaction where ambiguity vanishes due to periodicity, the scaling factor $\frac{1}{MN}$ preserves the ambiguity component as a distinguishable residual phase:
    \begin{equation}\label{eq:phi_explicit}
        \phi(l_{p,u}, k_i) = 2\pi \cdot l_{p,u} \frac{k_i}{MN} = \underbrace{2\pi \frac{\tilde{k}_i l_{p,u}}{MN}}_{\text{Base Phase}} \times \underbrace{2\pi \frac{N_{\mathrm{amb}, i} l_{p,u}}{M}}_{\text{Ambiguity Signature}}.
    \end{equation}
    This explicit dependence on $N_{\mathrm{amb}, i}$, scaled by the pilot's delay index $l_{p,u}$, provides the unique signature required for ambiguity estimation.

    With the phase rotation established, we now apply the remaining channel effects—specifically the delay-domain interaction and the Doppler-domain spreading—to the pilot signal. The complete received signal structure for the $u$-th pilot via the $i$-th path is expressed as:
    \begin{equation}\label{eq:y_pilot_matrix_form}
        \mathbf{Y}_{\mathrm{p}, u}^{(i)} = h_i x_{p,u} e^{j\phi(l_{p,u}, k_i)} \underbrace{\left[ \mathbf{\Pi}_{M,l_i} \mathbf{e}_{l_{p,u}}^{(M)} \right]}_{\text{Delay shift}}
        \underbrace{\left[ \left((\mathbf{e}_{k_{p,u}}^{(N)})^T
        \frac{1}{N} \sum_n \beta_n \mathbf{\Pi}_{N,n}^T \right)  \right]}_{\text{Doppler spreading}}
    \end{equation}
    where $\beta_n$, the fractional Doppler spreading coefficients defined in \cite[Eq.(16)]{xia2024achieving}, exhibit a maximum value at $k=K_i$.

    The first bracketed term in \eqref{eq:y_pilot_matrix_form} corresponds to the circular delay shift, moving the pilot to the index $\langle l_{p,u} + l_i\rangle_M$. The second term captures the spreading effect across the Doppler dimension. Evaluating these matrix operations leads to the final peak response at the receiver grid ${\langle l_i + l_{p,u} \rangle_M, \langle \tilde{K}_i + k_{p,u} \rangle_N}$:
    \begin{align}\label{eq:y_pilot_final}
        {y}_{p,u}^{(i)}=[\mathbf{Y}_{\mathrm{p}, u}^{(i)}]_{\langle l_i + l_{p,u} \rangle_M, \langle \tilde{K}_i + k_{p,u} \rangle_N} \approx h_i \lambda_{i,u} x_{p,u} \cdot e^{j \phi(l_{p,u}, k_i)} S(\kappa_i),
    \end{align}
    where $S(\kappa_i) \triangleq \frac{1}{N} e^{j \pi \kappa_i (N-1)/N} \mathrm{asinc}_N(\kappa_i)$ represents the Doppler spreading effect, with $\kappa_i$ denoting the fractional part of the Doppler shift.

    However, identifying the peak location alone is insufficient. As indicated in \eqref{eq:y_pilot_final}, the phase of the detected pilot signal is corrupted by the unknown channel phase $\angle h_i$ and the phase rotation induced by the fractional Doppler term $S(\kappa_i)$. These unknown parameters prevent the direct extraction of the Doppler ambiguity from a single pilot observation. To overcome this limitation, we propose a multi-pilot ambiguity estimation scheme in Section \ref{subsec:ambiguity_estimation}, which exploits the phase relationships between specific pilots to isolate the ambiguity term.

    In practice, the exact path delay $l_i$ and the wrapped integer Doppler shift $\tilde{K}_i$ are unknown. Nevertheless, since the relative configuration of the pilot symbols is fixed, we can exploit the combined energy of all pilots to locate the path. To achieve this, we employ a non-coherent energy accumulation scheme on the discrete DD grid. The coarse estimates $(\hat{l}_{i}, \hat{\tilde{K}}_{i})$ for the $i$-th path are determined by maximizing the summed magnitude:
    \begin{equation}\label{eq:coarse_estimation}
        (\hat{l}_{i}, \hat{\tilde{K}}_i) = \arg \max_{l, k} \sum_{u=1}^{N_{\mathrm{p}}} \left| \left( \mathbf{Y}_{\text{DD}}^{(i)} \right)_{\langle l + l_{p,u} \rangle_M, \langle k + k_{p,u} \rangle_N} \right|,
    \end{equation}
    where $\mathbf{Y}_{\text{DD}}^{(i)}$ denotes the residual of received signal at the $i$-th iteration.

    Once the coarse path parameters are identified, the complex value of the $u$-th received pilot signal is extracted from the global received grid based on the estimated coordinates:
    \begin{equation}
        \hat{y}_{p,u}^{(i)} = \left[ \mathbf{Y}_{\text{DD}}^{(i)} \right]_{\langle \hat{l}_i + l_{p,u} \rangle_M, \langle \hat{\tilde{K}}_i + k_{p,u} \rangle_N}.
    \end{equation}
    These extracted observations $\hat{y}_{p,u}^{(i)}$ serve as the input for the Doppler ambiguity estimation algorithm detailed in Section \ref{subsec:ambiguity_estimation}.

    \subsection{Proposed Pairwise Difference and Voting Algorithm}
    \label{subsec:ambiguity_estimation}

    Building upon the signal model derived in \eqref{eq:y_pilot_final}, we propose a robust ambiguity estimation algorithm based on differential pilot processing. To eliminate the nuisance parameters identified in the previous subsection, we exploit the fact that these terms are common across all pilots associated with the same propagation path. Consequently, by computing the conjugate product (phase difference) between selected pilot pairs, these unknown factors cancel out, effectively isolating the phase term containing the desired Doppler ambiguity.

    \subsubsection{Pairwise Differential Processing}
    Consider two distinct pilots $u$ and $u'$ ($u \neq u'$) associated with the $i$-th path, detected at the peak DD coordinates $(l_{i,u}, k_{i,u})$ and $(l_{i,u'}, k_{i,u'})$, respectively. Using \eqref{eq:y_pilot_final}, the ratio of their peak received signals is expressed as:
    \begin{align}
        \mathcal{R}_{u',u} \triangleq \frac{{y}_{p,u'}^{(i)}}{{y}_{p,u}^{(i)}}
        &= \frac{x_{p,u'}}{x_{p,u}} e^{j \left( \phi(l_{p,u'}, k_i) - \phi(l_{p,u}, k_i) \right)}
    \end{align}

    It is evident that the unknown complex channel gain $h_i$ and the spreading term $S(\kappa_i)$ cancel out perfectly. The phase of the ratio $\mathcal{R}_{u',u}$ becomes a linear function of the total normalized Doppler shift $k_i$, scaled by the known pilot spacing:
    \begin{equation}
        \angle \left( \mathcal{R}_{u',u} \right) = \phi(l_{p,u'}, k_i) - \phi(l_{p,u}, k_i) = \frac{2\pi k_i}{MN} (l_{p,u'} - l_{p,u}).
    \end{equation}
    Let $\Delta l_{u',u} = l_{p,u'} - l_{p,u}$ denote the delay spacing between pilot $u'$ and pilot $u$. We can now obtain an estimate of the total normalized Doppler shift, denoted as $\hat{k}_{est}^{(u',u)}$, by inverting the phase difference:
    \begin{equation}\label{eq:k_est_pair}
        \hat{k}_{i}^{(u',u)} = \frac{MN}{2\pi \Delta l_{u',u}} \cdot \angle \left(\mathcal{R}_{u',u}\right).
    \end{equation}

    \subsubsection{Ambiguity Resolution and Majority Voting}
    The estimated total Doppler $\hat{k}_{est}^{(u',u)}$ is theoretically equal to the true Doppler $k_i = \tilde{k}_i + N_{\mathrm{amb},i}N$. Since the base Doppler index $\tilde{k}_i$ (wrapped within $[-N/2, N/2)$) is readily observable from the peak position on the DD grid, the integer ambiguity for the pair $(u',u)$ can be isolated as:
    \begin{equation}
        \hat{N}_{\mathrm{amb,i}}^{(u',u)} = \mathrm{round}\left( \frac{\hat{k}_{i}^{(u',u)} - \hat{{\tilde{K}}}_i}{N} \right).
    \end{equation}

    To mitigate the impact of noise and potential estimation errors from small phase perturbations, we perform this estimation for all available pilot pairs. For a system with $N_{\mathrm{p}}$ pilots, there are $\binom{N_{\mathrm{p}}}{2}$ unique pairs. We employ a majority voting strategy to determine the final ambiguity integer:
    \begin{equation}
        \hat{N}_{\mathrm{amb,i}} = \mathop{\mathrm{mode}}_{u' > u} \left\{ \hat{N}_{\mathrm{amb,i}}^{(u',u)} \right\}.
    \end{equation}
    This voting mechanism significantly enhances the robustness of the estimator, ensuring reliable Doppler recovery even under low SNR conditions.

    It is worth noting that the computational complexity of this proposed ambiguity estimation scheme is negligible compared to the subsequent MLE-based channel estimation. The voting process relies solely on simple algebraic operations and sorting, imposing minimal overhead. Crucially, by pre-determining $\hat{N}_{\mathrm{amb}, i}$, the search space for the subsequent MLE stage is strictly confined to the fractional Doppler region (i.e., the standard ambiguity-free range). This effectively eliminates the need for a prohibitive brute-force search over all possible ambiguity integers, ensuring that high-precision estimation does not incur additional computational cost relative to a conventional ambiguity-free scenario.

    \subsection{Parametric Channel Estimation and Data Detection}\label{subsec:channel_estimation}
    Once the Doppler ambiguity integer of $i$-th path $\hat{N}_{\mathrm{amb},i}$ is estimated, we perform the slow-time dimension phase compensation on the received DD domain signal. This is achieved by left-multiplying $\mathbf{Y}_{\mathrm{DD}}^{(i)}$ with the inverse phase rotation matrix $\mathbf{\Phi}(\hat{N}_{\mathrm{amb},i})^{-1}$:
    \begin{equation}
        \mathbf{Y}_{\mathrm{DD,com}}^{(i)}=
        \mathbf{\Phi}(\hat{N}_{\mathrm{amb,i}})^{-1}
        \mathbf{Y}_{\mathrm{DD}}^{(i)}
    \end{equation}

    Given the coarse integer estimates $(\hat{l}_{i}, \hat{K}_{i})$, we refine the Doppler shift to capture the fractional component $\kappa_i$. A local fine-resolution MLE search is performed within the neighborhood of $\hat{K}_{i}$:
    \begin{equation}
        \hat{\kappa}_{i}= \arg \max_{\kappa \in \Omega}
        \left|
        \text{vec}\{\mathbf{t}(\hat{l}_{i}, \hat{K}_{i}+\kappa)\}^H
        \text{vec}\{\mathbf{Y}_{\mathrm{DD,com}}^{(i)}\} \right|,
    \end{equation}
    where $\Omega=[-0.5, -0.5+\epsilon, \dots, 0.5]$ represents the fine search region with step size $\epsilon$, comprising a total of $N_{\kappa} = \frac{1}{\epsilon}+1$ search candidates. The term $\mathbf{t}(l,k)=\mathbf{Q}_{\mathrm{ISI}}^{\mathrm{Fast}}(l) \mathbf{D}(k) \tilde{\mathbf{X}}_{\mathrm{DD}}(l,k) \mathbf{Q}_{\mathrm{ISI}}^{\mathrm{Slow}}(k)$ denotes the correlation filter. The final Doppler estimate is given by $\hat{k}_{i}=\hat{K}_{i}+\hat{\kappa}_{i}+\hat{N}_{\mathrm{amb,i}}N$. Finally, the complex channel gain of the $i$-th path  is recovered by:
    \begin{equation}
        \hat{h}_{i} =
        \frac{\text{vec}\{\mathbf{t}(\hat{l}_{i}, \hat{k}_{i})\}^H
        \text{vec}\{ \mathbf{Y}_{\mathrm{DD}}^{(i)}\}}
        {\text{vec}\{\mathbf{t}(\hat{l}_{i}, \hat{k}_{i})\}^H\text{vec}\{\mathbf{t}(\hat{l}_{i}, \hat{k}_{i})\}}.
    \end{equation}

    Following the estimation of the $i$-th path, its contribution is reconstructed and subtracted from the received signal. This process is repeated iteratively to estimate the parameters for the subsequent strong paths.

    After estimating the parameters $\{\hat{h}_i, \hat{l}_i, \hat{k}_i\}$ for all $P$ paths, the effective channel matrix $\hat{\mathbf{H}}_{\text{DD}}$ is reconstructed. Finally, data detection is performed using the Message Passing Algorithm (MPA) described in \cite{raviteja2018interference}.

    \subsection{Computational Complexity Analysis}\label{subsec:complexity_analysis}

    We analyze the computational complexity of the proposed algorithm by quantifying the complex additions and multiplications required to estimate $P$ propagation paths.

    First, for the stage of coarse energy accumulation, summing the magnitudes of $N_{\mathrm{p}}$ pilot observations over the coarse search grid of size $l_{\max} \times N$ requires approximately $P\times N_{\mathrm{p}} \times N \times l_{\max}$ complex additions.
    Second, the Doppler ambiguity estimation stage resolves the integer ambiguity using $\binom{N_{\mathrm{p}}}{2}$ pairwise phase differences. This step incurs a cost of approximately $P \times N_{\mathrm{p}}^2$ complex multiplications. Given that $N_{\mathrm{p}}$ is typically small (e.g. $2 \sim 5$), this computational overhead is negligible.
    Finally, the MLE fine search constitutes the most resource-intensive stage. As detailed in \cite{xia2024achieving}, evaluating the likelihood function for a single DD candidate involves $2N$ $M$-point FFTs and $N$ $N$-point FFTs. Consequently, for $N_{\kappa}$ fractional candidates per path, this stage requires $P \times N_{\kappa} (MN^2 + N^2\log_2N + 2MN\log_2M)$ complex additions and $P \times N_{\kappa} ((3M +MN + l_{\max})N + \frac{N^2}{2}\log_2N + MN\log_2M)$ complex multiplications.

    To highlight the efficiency gain, we also analyze the computational complexity of the \textbf{exhaustive search-based} Extended MLE benchmark. Since Extended MLE lacks an explicit ambiguity estimation mechanism, it must resort to a brute-force search throughout the potential range of integer ambiguities $[-N_{\mathrm{amb}}^{\max}, N_{\mathrm{amb}}^{\max}]$ within the computationally intensive fine search stage. This expands the search space by a factor of $\eta = (2N_{\mathrm{amb}}^{\max} + 1)$. Consequently, the required operations increase to $P\times\eta\times N_{\kappa}(MN^2 + N^2\log_2N + 2MN\log_2M)$ complex additions and $P\times\eta\times N_{\kappa}((3M +MN + l_{\max})N + \frac{N^2}{2}\log_2N + MN\log_2M)$ complex multiplications.

    Furthermore, In practical high-mobility scenarios, the maximum ambiguity integer $N_{\mathrm{amb}}^{\max}$ is often unknown a priori, necessitating a conservatively large search window to avoid coverage loss. This results in a significant waste of computational resources on invalid candidates. In contrast, our proposed pilot-aided scheme pre-determines the unique integer ambiguity with negligible cost, thus bypassing the brute-force expansion factor $\eta$ and achieving a substantial reduction in overall complexity.

\section{Simulation Results}\label{simulation_result}

\begin{table}[htbp]
    \centering
    \caption{Simulation Parameters}
    \label{tab:parameters}
    \begin{tabularx}{\columnwidth}{X r}
        \toprule
        \textbf{Parameter} & \textbf{Value} \\
        \midrule
        Carrier Frequency, $f_c$ & 2 GHz \\
        Subcarrier Spacing, $\Delta f$ & 15 kHz \\
        Number of Subcarriers, $N$ & 32 \\
        Number of Slow-time Intervals, $M$ & 64 \\
        Number of Paths, $P$ & 4 \\
        Max. Normalized Delay Spread, $l_{\max}$ & 4 \\
        Max. Normalized Doppler Spread, $k_{\max}$ & 79 \\
        Max. Doppler Ambiguity Integer, $N_{\mathrm{amb}}^{\max}$ & 3 \\
        MLE Fine Search Step, $\epsilon$ & 0.1 \\
        Modulation Scheme & QPSK \\
        \bottomrule
    \end{tabularx}
\end{table}

In this section, we evaluate the performance of the proposed pilot-aided Doppler ambiguity and channel estimation schemes based on an OTFS system operating at $2$ GHz over a sparse multi-path channel ($P=4$).
The random delay indices are generated in $[0, l_{\max}]$, and the random Doppler shifts are generated in $[-k_{\max}, k_{\max}]$.
To simulate the extreme mobility of LEO satellites, we set $k_{\max}=79$, which corresponds to a maximum relative velocity of $v_{\max} = 20,000$ km/h.
In our simulations, we employ practical rectangular transmit and receive pulse shaping waveforms.
The signal-to-noise ratio is defined as $\text{SNR}_{\text{data}} = \mathbb{E}_{l,k}[|d_{l,k}|^2]/\sigma^2_n$, where $\sigma^2_n$ is the AWGN variance.
The Pilot-to-Data Power Ratio is defined as $\text{PDR} = \mathbb{E}_{u}[|x_{p,u}|^2] / \mathbb{E}_{l,k}[|d_{l,k}|^2]$.
In terms of pilot configuration, we set $N_p=2$ for the EP-GZ scheme to limit overhead, whereas for the DSP scheme, we adopt $N_p=5$ to improve the robustness of the estimation.
The pilot delay spacing is equal to the maximum delay spread $l_{\max}$.
All key parameters are summarized in Table \ref{tab:parameters}.

It is worth noting that while larger subcarrier spacings (e.g. 60 kHz) are often used to mitigate Doppler effects, we deliberately adopt $\Delta f = 15$ kHz in this paper. This setup represents a \textit{stress-test scenario} prioritizing spectral efficiency, where Doppler ambiguity is most severe (as discussed in the footnote), thereby rigorously validating the robustness of our proposed estimator.

\subsection{Doppler Ambiguity Detection Performance}\label{subsec:dadp}

\begin{figure}[!t]
    \graphicspath{{./Figure/}}
    \centering
    \includegraphics[scale=0.65]{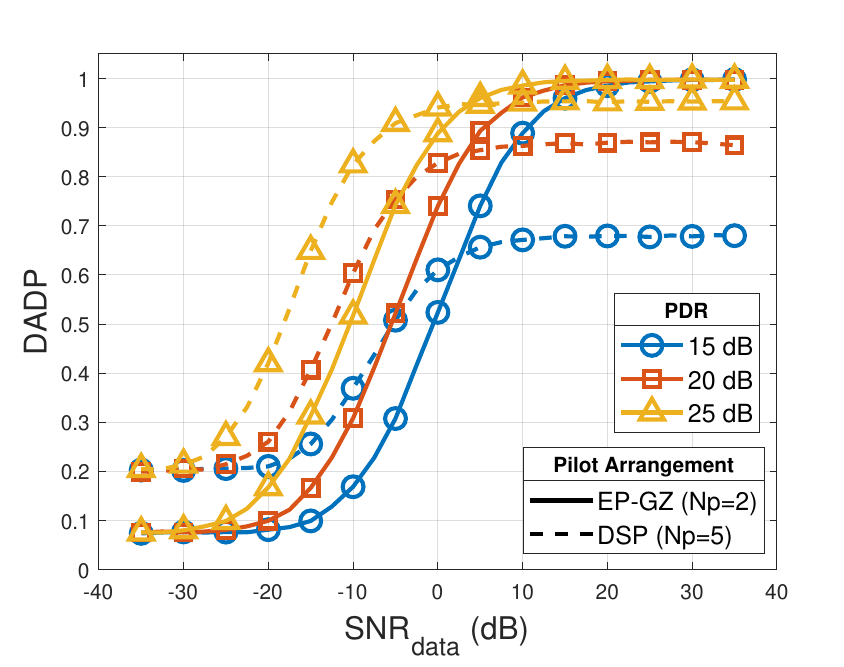}
    \caption{Average Doppler Ambiguity Detection Probability for different PDR and pilot arrangement.}
    \label{avg_dadp}
\end{figure}

Fig. \ref{avg_dadp} compares the average Doppler Ambiguity Detection Probability (DADP) versus SNR for EP-GZ and DSP schemes under different Pilot-to-Data Power Ratios (PDR). The average DADP is defined as the ratio of the number of successfully resolved ambiguity integers to the total number of paths.

For the EP-GZ scheme, the inclusion of a guard zone effectively isolates the pilot from data interference. Consequently, as long as the SNR is sufficiently high, the DADP converges to approximately $1$, showing robustness across different PDR settings. In contrast, the DSP scheme lacks a guard zone and is therefore susceptible to interference from data symbols, necessitating a higher PDR to mitigate this effect. In particular, the detection probability for the DSP scheme does not strictly converge to $1$. This is attributed to the existence of extremely weak propagation paths in the simulation; the pilot signals associated with these weak paths are overwhelmed by severe data interference originating from stronger paths. However, since these undetected weak paths contribute minimally to the total signal energy, they can be treated as residual noise. As a result, their impact on overall system performance is negligible, as will be demonstrated by the BER results in Section \ref{subsec:system_performance}.

\begin{figure*}[ht]
    \centering
    \setkeys{Gin}{width=0.32\linewidth} 

    \subfloat[BER vs. $\text{SNR}_{\text{data}}$.  \label{fig:ber_pdr25}]{\includegraphics{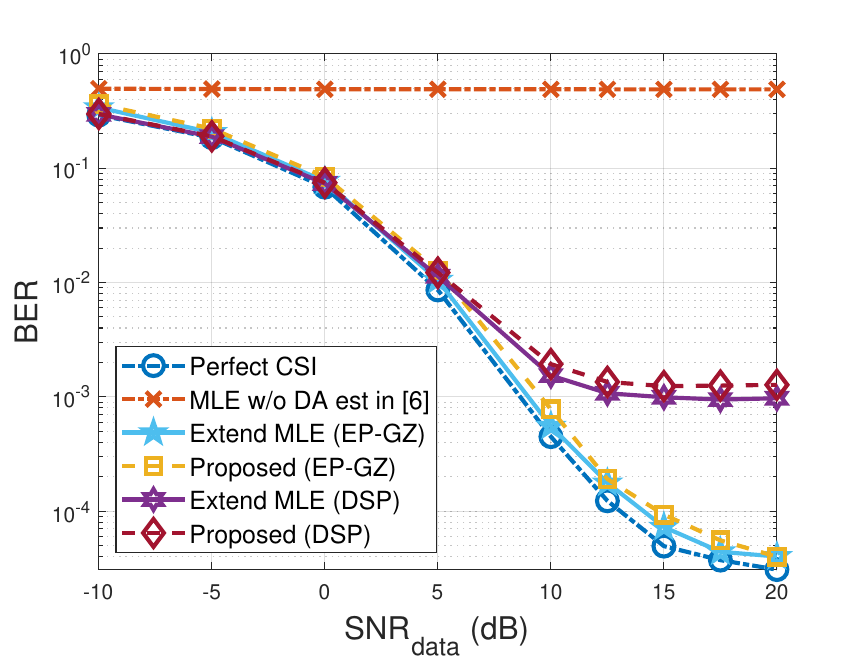} }\hfil
    \subfloat[NMSE vs. $\text{SNR}_{\text{data}}$. \label{fig:nmse_pdr25}]{\includegraphics{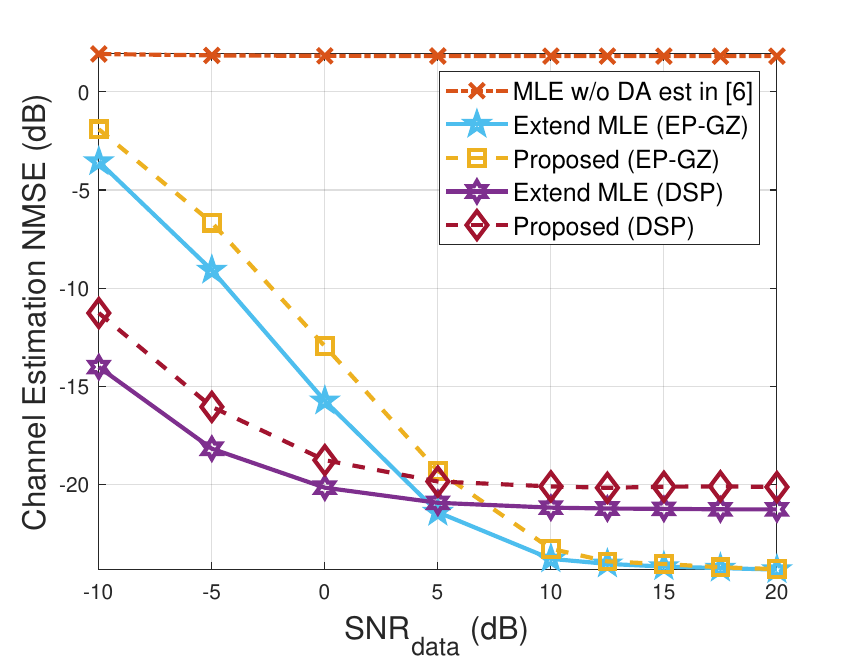} }\hfil
    \subfloat[ESE vs. $\text{SNR}_{\text{data}}$.  \label{fig:se_pdr25}]{\includegraphics{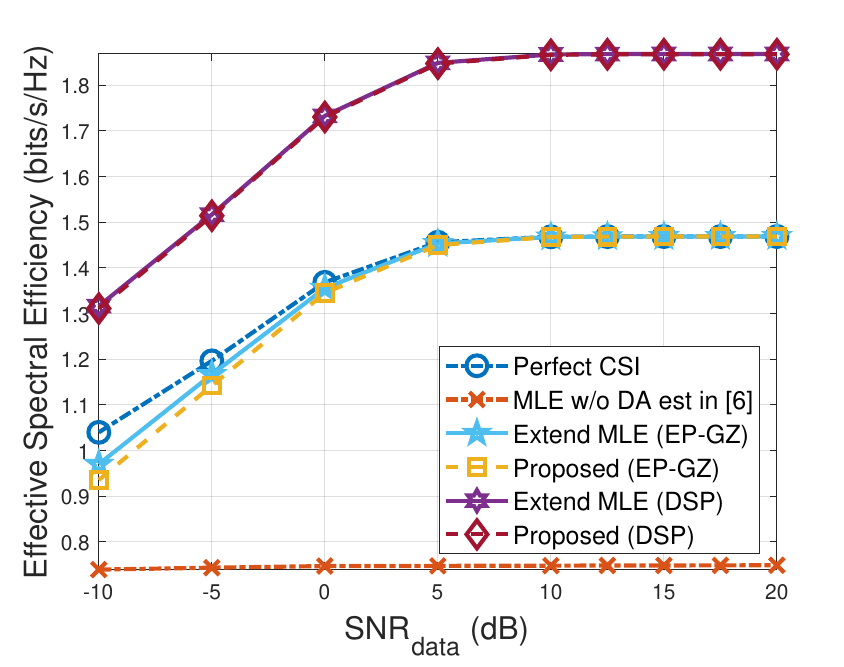} }

\caption{System simulation results with PDR = 25 dB: (a) Bit Error Rate, (b) Channel estimation NMSE, and (c) Effective Spectral Efficiency.}
\label{fig:overall_results}
\end{figure*}

\subsection{System Performance}\label{subsec:system_performance}
In the following performance evaluation, we adopt a relatively high PDR of 25 dB. This value is specifically chosen to ensure robust ambiguity resolution for the DSP scheme, which operates in an interference-limited regime due to the superposition of data symbols. In contrast, the EP-GZ scheme, benefiting from an interference-free guard zone (noise-limited regime), can achieve reliable detection at much lower PDR levels. However, to provide a fair comparison of the achievable performance upper bounds, the same PDR is maintained for both schemes.

Fig. \ref{fig:overall_results} presents the BER, NMSE, and ESE performance at PDR = 25 dB. We compare the proposed scheme against three benchmarks: 1) \textbf{Perfect CSI} (ideal parameter knowledge); 2) \textbf{Standard MLE} \cite{khan2023low} (ignoring integer ambiguity); and 3) \textbf{Extended MLE} (exhaustive search upper bound).

Figs. \ref{fig:ber_pdr25} and \ref{fig:nmse_pdr25} reveal distinct performance characteristics among these methods. The method in \cite{khan2023low} suffers from catastrophic degradation, with the BER hovering around $0.5$. This failure arises because neglecting the Doppler ambiguity introduces significant uncompensated phase shifts, leading to a severe model mismatch. Consequently, the channel estimation becomes erroneous, making data detection impossible. In contrast, the Extended MLE scheme successfully recovers the channel parameters and matches the performance of the Perfect CSI case. However, this comes at the cost of prohibitive computational complexity, as it necessitates an exhaustive search over a potentially large and unknown integer ambiguity space, causing substantial resource wastage in practical systems. Most importantly, the proposed scheme achieves a performance closely approaching that of the Extended MLE and Perfect CSI benchmarks. By accurately resolving the Doppler ambiguity prior to channel estimation, the proposed method effectively eliminates the model mismatch while maintaining computational complexity similar to the standard MLE. This shows that our low-complexity strategy successfully reconciles the trade-off between estimation accuracy and computational efficiency.

Comparing the two pilot arrangements, the EP-GZ scheme exhibits superior performance in terms of BER and NMSE. This is attributed to the presence of the guard zone, which creates an interference-free observation window for the pilot, allowing the estimation error to decay continuously as SNR increases. In contrast, the DSP scheme is subject to mutual interference from the data symbols. As a result, it encounters an error floor in the high SNR regime, preventing the NMSE and BER from converging as deeply as in the EP-GZ case. Nevertheless, it is crucial to emphasize that regardless of the specific pilot configuration, the proposed framework consistently delivers robust performance, effectively approaching the performance upper bound established by the Extended MLE benchmark.

Finally, Fig. \ref{fig:se_pdr25} illustrates the trade-off regarding Effective Spectral Efficiency (ESE). It is observed that the DSP scheme achieves a higher ESE compared to the EP-GZ scheme. This is because the DSP configuration eliminates the overhead associated with guard zones, allowing a larger number of data symbols to be transmitted. Consequently, while the EP-GZ scheme provides superior interference suppression (as evidenced by the lower BER and NMSE), the DSP scheme offers a competitive advantage in spectral efficiency.

\section{Conclusion}\label{ch5}
In this paper, we address the critical challenge of integer Doppler ambiguity in high-frequency OTFS systems. We demonstrated that this ambiguity induces a structured phase rotation along the delay dimension, causing conventional parametric estimators to fail. To resolve this, we proposed a low-complexity pilot-aided framework that efficiently identifies the ambiguity integer through pairwise pilot phase differences. Simulation results verify that the proposed estimator eliminates the error floor, achieving performance comparable to the high complexity exhaustive search benchmark. Ultimately, by overcoming the conventional Doppler limitation, our approach extends the operational envelope of OTFS, solidifying its position as a promising candidate for next-generation high-mobility communications.

\bibliography{ref}
\bibliographystyle{IEEEtran}
\end{document}